\documentclass[sigconf]{acmart}
\PassOptionsToPackage{bookmarks=false}{hyperref}
\usepackage{booktabs} % For formal tables
\usepackage[toc,page]{appendix} % for appendix

\usepackage{paralist}
\usepackage{etoolbox}
\usepackage{microtype}
\DeclareCaptionLabelFormat{andtable}{#1~#2  \&  \tablename~\thetable}

\usepackage[bookmarks=false]{hyperref}
\usepackage{fancybox}
\usepackage{graphicx}
\graphicspath{{Figures/}}
\DeclareGraphicsExtensions{.pdf,.jpeg,.png}
\usepackage{amsmath}
\usepackage{subfigure}
\usepackage{caption}
\usepackage{fancyvrb}
\usepackage{color}
\usepackage{algorithmic}
\PassOptionsToPackage{hyphens}{url}\usepackage{hyperref}
\usepackage{array}
\usepackage{mdwmath}
\usepackage{mdwtab}
\usepackage{booktabs}
\usepackage{listings}
\usepackage{eqparbox}
\usepackage{stfloats}
\usepackage{comment}
\usepackage{framed}
\usepackage{ifthen}
\usepackage{balance}
\usepackage{enumitem}
\usepackage{balance}
\usepackage[turnoff]{notes}

\usepackage{multirow}
\usepackage{xspace}
\usepackage{listings}
\newcommand{\rom}[1]{\uppercase\expandafter{\romannumeral #1\relax}}

\newcommand{\etal}{\hbox{{et al.}}\xspace}
\newcommand{\eg}{\hbox{{e.g.,}}\xspace}
\newcommand{\ie}{\hbox{{i.e.}}\xspace}

\newcommand{\wrt}{\hbox{{w.r.t.}}\xspace}

\newcommand{\etc}{\hbox{{etc.}}\xspace}

\definecolor{gray50}{gray}{.5}
\definecolor{gray40}{gray}{.6}
\definecolor{gray30}{gray}{.7}
\definecolor{gray20}{gray}{.8}
\definecolor{gray10}{gray}{.9}
\definecolor{gray05}{gray}{.95}

\newlength\Linewidth
\def\findlength{\setlength\Linewidth\linewidth
\addtolength\Linewidth{-4\fboxrule}
\addtolength\Linewidth{-3\fboxsep}
}
\newenvironment{examplebox}{\par\begingroup
   \setlength{\fboxsep}{5pt}\findlength
   \setbox0=\vbox\bgroup\noindent
   \hsize=0.95\linewidth
   \begin{minipage}{0.95\linewidth}\normalsize}
    {\end{minipage}\egroup
    \textcolor{gray20}{\fboxsep1.5pt\fbox
     {\fboxsep2pt\colorbox{gray05}{\normalcolor\box0}}} %this is width of the RQ box
    \endgroup\par\noindent
    \normalcolor\ignorespacesafterend}

\newcounter{RQCounter}
\newcounter{RQACounter}

\newcommand{\RQ}[2]{%
\refstepcounter{RQCounter} \label{#1}
 \begin{center}	
  \begin{examplebox}
   \textbf{RQ\arabic{RQCounter}.}~#2
  \end{examplebox}	 
 \end{center}
}

\newcommand{\RQA}[2]{%
\refstepcounter{RQACounter} \label{#1}
\vspace{0.1in} \noindent\textbf{RQ\arabic{RQACounter}.~#2 \vspace{0.05in}}

}

\usepackage{tcolorbox}% http://ctan.org/pkg/tcolorbox
\definecolor{mycolor}{rgb}{0.122, 0.435, 0.698}% Rule colour
\newcommand{\mybox}[1]{%
  \setbox0=\hbox{#1}%
  \setlength{\@tempdima}{\dimexpr\wd0+13pt}%
  \begin{tcolorbox}[colframe=mycolor,boxrule=0.5pt,arc=4pt,
      left=6pt,right=6pt,top=6pt,bottom=6pt,boxsep=0pt,width=\@tempdima]
    #1
  \end{tcolorbox}
}

\newcommand{\RS}[2]{%
\begin{tcolorbox}[colback=white,boxrule=0.5pt,arc=4pt]
\textbf{\underline{Result {\ref{#1}}}:~}{{\em #2}}%
\end{tcolorbox}
}

\definecolor{javared}{rgb}{0.6,0,0} % for strings
\definecolor{javagreen}{rgb}{0.25,0.5,0.35} % comments
\definecolor{javapurple}{rgb}{0.5,0,0.35} % keywords
\definecolor{javadocblue}{rgb}{0.25,0.35,0.75} % javadoc

\lstdefinestyle{customc}{
  belowcaptionskip=\baselineskip,
  breaklines=true,
  xleftmargin=\parindent,
  language=java,
  showstringspaces=false,
  basicstyle=\scriptsize\ttfamily,
  keywordstyle=\bfseries\color{javapurple},
  commentstyle=\itshape\blue,
  belowskip=-10pt,
  aboveskip=-5pt
}

\DeclareCaptionLabelFormat{andimage}{#1~#2  \&  \figurename~\thefigure}

\newcommand{\cscore}{\textit{codeness score}\xspace}

\newcommand{\gh}{GitHub\xspace}
\newcommand{\md}[1]{{\scriptsize \todo{Masud:  {\color{blue} #1}}}}

\newcommand{\GPSE}{{\sc GPSE}\xspace}
\newcommand\blue[1]{\textcolor[rgb]{0.00,0.00,1.00}{{#1}}}

\usepackage{epstopdf}
\lstdefinestyle{example} {
	frame=tb,
	basicstyle=\footnotesize\ttfamily,
	numbers=left,
	language=C
}

\begin{document}
% Copyright, text after fillup acm copyright form
\copyrightyear{2018} 
\acmYear{2018} 
\setcopyright{acmcopyright}
\acmConference[MSR '18]{MSR '18: 15th International Conference on Mining Software Repositories }{May 28--29, 2018}{Gothenburg, Sweden}
%\acmBooktitle{MSR '18: MSR '18: 15th International Conference on Mining Software Repositories , May 28--29, 2018, Gothenburg, Sweden}
\acmPrice{15.00}
\acmDOI{10.1145/3196398.3196425}
\acmISBN{978-1-4503-5716-6/18/05}

\title{Evaluating How Developers Use General-Purpose Web-Search for Code Retrieval}

\author{Md Masudur Rahman}
%\authornote{any note about this author.}
%\orcid{author orcid}
\affiliation{%
  \institution{University of Virginia}
}
\email{masud@virginia.edu}

\author{Jed Barson}
\affiliation{%
  \institution{University of Virginia}
}
\email{jb3bt@virginia.edu}

\author{Sydney Paul}
\affiliation{%
  \institution{Clemson University}
}
\email{sepaul@g.clemson.edu}

\author{Joshua Kayani}
\affiliation{%
  \institution{North Carolina State University}
}
\email{jkayani@ncsu.edu}

\author{Federico Andr\'es Lois}
\affiliation{%
  \institution{Codealike, Argentina}
}
\email{Federico.lois@corvalius.com}

\author{Sebasti\'an Fernandez Quezada}
\affiliation{%
   \institution{Codealike, Argentina} 
}
\email{sebastian.quezada@corvalius.com}

\author{Christopher Parnin}
\affiliation{%
  \institution{North Carolina State University}
}
\email{cjparnin@ncsu.edu}

\author{Kathryn T. Stolee}
\affiliation{%
  \institution{North Carolina State University}
}
\email{ktstolee@ncsu.edu}

\author{Baishakhi Ray}
%\authornote{any note about this author.}
%\orcid{author orcid}
\affiliation{%
  \institution{University of Virginia}
}
\email{rayb@virginia.edu}

\begin{abstract}
Search is an integral part of a software development process. Developers often use search engines to look for information during development, including reusable code snippets, API understanding, and reference examples. Developers tend to prefer general-purpose search engines like Google, which are often not optimized for code related documents and use search strategies and ranking techniques that are more optimized for generic, non-code related information.

In this paper, we explore whether a general purpose search engine like Google is an optimal choice for code-related searches. In particular, we investigate whether 
%(i) users' search behaviors change while using Google for code search, and (ii) 
the performance of searching with Google varies for code vs. non-code related searches. To analyze this, we collect search logs from 310 developers that contains nearly 150,000 search queries from Google and the associated result clicks. 
To differentiate between code-related searches and non-code-related searches, we build a model which identifies the code intent of queries. Leveraging this model, we build an automatic classifier that detects a code and non-code related query.
%by analyzing the intent of its individual token. 
We confirm the effectiveness of the classifier on manually annotated queries where the classifier achieves a \textit{precision} of $87\%$, a \textit{recall} of $86\%$, and an \textit{F1-score} of $87\%$. 
We apply this classifier to automatically annotate all the queries in the dataset. Analyzing this dataset, we observe that code related searching often requires more effort (\eg~time, result clicks, and query modifications) than general non-code search, which indicates code search performance with a general search engine is less effective.

%In this paper, we explore how the search behavior of users and performance of the general-purpose search engine (\ie Google) vary for code and non-code related search. To analyze this, we build a classifier which identifies a code related query by analyzing the intent of its individual token. We observe that code related search often required more effort than general non-code search which indicates code search performance in general search engine is less effective. ~\todo{more results?}

\end{abstract}



% \keywords{ACM proceedings, \LaTeX, text tagging}
\renewcommand{\shortauthors}{Rahman et al.}

\maketitle

%!Tex root=main.tex

\section{Introduction}
\label{sec:intro}	

Search plays an important role in fulfilling users' information needs. In particular, search for code has been an integral part of software development processes in the past ~\cite{bauer2014exploratory, umarji2008archetypal, sadowski2015developers, Xia2017}.
Developers often use a search engine for various information needs, including finding reusable code snippets, understanding APIs, locating reference examples,  learning unfamiliar concepts, remembering syntactic details, identifying appropriate third-party libraries, and debugging ~\cite{hucka2016software,sadowski2015developers, Xia2017}. In the  literature, code search has been studied extensively and researchers proposed many approaches to improve code search performance~\cite{raghothaman2016swim, niu2017learning, martie2017understanding, bajracharya2006sourcerer, lazzarini2007codegenie, reiss2009semantics, sindhgatta2006using, thummalapenta2007parseweb, durao2008applying,haiduc2013automatic, holmes2005using, lemos2014thesaurus, linstead2009sourcerer, mcmillan2011portfolio, ye2002supporting}. 
%~\todo{add more if related}.

%about research domain
%Due to the increasing demand of code search in software development, many dedicated commercial code search engine has been developed (\ie ~\cite{krugle-search, searchcode-search, koder-openhub, google-code-search-dep}) and unfortunately, many of them (\ie ~\cite{koder-openhub, google-code-search-dep}) are now deprecated

In practice, to support the increasing need for code search in software development, several commercial search engines have been developed, such as Google Code Search~\cite{google-code-search-dep},  Black Duck Open Hub Code Search~\cite{koder-openhub}, and others (\eg~\cite{krugle-search, searchcode-search}). Unfortunately, many of them (\eg ~\cite{koder-openhub, google-code-search-dep}) are now obsolete.
%which left few alternatives for developers to search code on Web. 
%Because of the suitable alternative developers often use general purpose search engine for code related queries. 
Thus, programmers tend to turn to a general purpose search engine  (\eg Google~\cite{google-search}, Yahoo~\cite{yahoo-search}, Bing~\cite{bing-search}) to search for code~\cite{sim2011well, hucka2016software, stolee2014tosem}, and software ~\cite{hucka2016software}, and they rarely use a dedicated code search engines~\cite{hucka2016software}. Among several general-purpose search engines, \textit{Google} has been found to be the most frequently used search engine for software development related searches~\cite{sim2011well}. 

These general purpose search engines (\GPSE) are usually optimized for textual search~\cite{hucka2016software} and treat code as plain text when used for searching code. Thus, they tend to ignore the underlying semantics of the code. In fact, Google's dedicated code search engine~\cite{google-code-search-article} used an additional layer of n-gram based regular expression matching technique to cater the special needs of code search. Using \GPSE for code search might be a reason that despite the tremendous increase in online resources (\eg \gh, SourceForge, StackOverflow, API documentation), suitably locating reusable source code still remains a major challenge\textemdash developers often fail to locate the intended code using different search approaches including web-search~\cite{hucka2016software}. Surveys have shown that developers look at an average of 3.5 snippets of code before finding something useful for their task at hand~\cite{stolee2014tosem}.

 Despite their limitations, {\sc GPSE}s are the most popular choice for code search and will continue to be like that, in all likelihood, because they are lightweight, easy to use, and have sophisticated web interfaces~\cite{martie2017understanding}. Thus, it is worthwhile to evaluate \textit{how {\sc GPSE}s perform while used for code vs.~general non-code related search} so that we can better understand \GPSE's shortcomings in code domain and tune them accordingly. 
In this paper, we shed some empirical light to explore this question by studying Google search~\cite{google-search}. 
In particular, we investigate how the search behavior of users and performance of the search engine vary for code related searches compared to non-code related searches. 

To this end, we analyzed $14$ months of web search logs from $310$ developers using a Google Chrome plugin, which tracks browsing activities of its users~\cite{codealike}. In total, we analyzed $149,610$ Google search queries. Since these are web search logs of the developers during their working time, the logs contain both code and non-code related queries, although they are  not annotated as such. First, we develop an automated technique to classify these queries to code vs.~non-code. We leverage Stack Overflow~\cite{stackoverflow} tags to extract code-related tokens. A {\em codeness} score is calculated for each query based on how many stack-overflow tokens are present in it. A higher codeness score indicates the query is more likely to be code related. 
A manual evaluation shows that the classifier achieves a \textit{precision} of $87\%$, a \textit{recall} of $86\%$ to successfully classify code vs.~non-code queries. Using this classifier we find 88,577 ($59.21\%$) code related queries and 61,033 ($40.79\%$) non-code queries in our dataset.
We use this annotated data to analyze the differences between code and non-code related search for \GPSE. We study both query characteristics (RQ1) and developers' effort (RQ2 \& RQ3) and find that: %developers have to put more effort for code search than non-code search. In particular, 

\begin{enumerate}[leftmargin=*,topsep=1pt]
    \item 
    A single code query is, in general, larger and uses a smaller vocabulary than a non-code query (see RQ1).
    \item 
    To retrieve the intended answer, users have to spend more time on a single code query and have to modify the queries more often than the non-code queries  (see RQ~\ref{rq2}).
    \item 
    To complete a code related search task, users require more queries, more URLs clicked, and overall more time than non-code related search tasks  (see RQ~\ref{rq3}).
    
\end{enumerate}

% the former often requires more keywords (details in RQ~\ref{rq1}), users spend more time and query modification) when they search for code related issues compared to non-code (details in RQ~\ref{rq2}).
% Similarly, to complete a code related search task, users require more queries, more URLs clicked, more time than non-code related search tasks, and users often show more patience when they look for code-related information (details in RQ~\ref{rq3}).

Several empirical studies have been performed to identify how developers search for code~\cite{sadowski2015developers}, what type of code issues they search on Web~\cite{Xia2017}, and how the performance varies when developers search for code with different search engines~\cite{sim2011well}. 
Yet, it remains less explored how code search is different than searching for general information, \ie non-code search. Little is known about how the \GPSE  performs on code search compared to others. 
In this paper, we seek to answer these questions. In summary, we make the following contributions: 

\begin{itemize}[leftmargin=*]
    \item Build and evaluate a novel technique to automatically % which can estimate the code intent of a query and 
    classify a search query to code vs.~non-code %based on  effectively 
    (Section~\ref{query-classification}).
    \item Analyze the query characteristics and how it differs between code vs.~non-code queries in general-purpose web search (RQ~\ref{rq1}).
    \item Analyze users' effort in retrieving the intended result for code and non-code related queries (RQ~\ref{rq2} and RQ~\ref{3}).
    %\item Analyze how query (RQ~\ref{rq2}) and task (RQ~\ref{rq3}) search behavior differs for code and non-code related queries
\end{itemize}

%paper organization
%\todo{change with new RQ formating}
We organize the rest of the paper as follows. We start by describing background information and research questions in Section~\ref{sec:back}. Then we discuss our code intent model in Section~\ref{codeness-model}. We discuss our methodology in details including data collection, query extraction and annotation, and classifier evaluation in Section~\ref{sec:method}. After that, we analyze our experimental results in Section~\ref{sec:exp}. We discuss the implication of our code intent model and findings in Section~\ref{sec:dicussion}. Then we discuss related work in Section~\ref{sec:related}, possible threats to validity in Section~\ref{sec:threats}, and conclude in Section~\ref{sec:con}. %~\todo{update if discussion section added}

\section{Research Questions}\label{sec:back}
Typically, code related artifacts (\eg source code, bug reports,  API documentation, \etc) are different from general documents, such as news, Wikipedia articles, or other non-code information sources~\cite{hellendoorn2017deep}. While the latter is primarily composed of natural languages, the code related documents can be a mix of programming and natural languages. However,  \GPSE treats source code as text and ignores all the programming language related features. For example, the source code is {\em less ambiguous} than natural language so that the code can be interpreted by a compiler. However, \GPSE ignores the syntactic and semantic features of the source code and thus, cannot interpret the underlying behavior. Thus, using \GPSE, locating similar code or retrieving code examples becomes difficult unless both query and the documents use similar vocabulary~\cite{stolee2014tosem}. But, since source code contains {\em open vocabulary} (\ie  developers can coin new variable names without changing the semantics of the programs), searching for code somewhat becomes a guessing game for \GPSE.  

%Users search for many reasons, which include code related search in a general-purpose search engine. 
In this paper, we empirically evaluate the impact of using \GPSE for code related search. In particular, we investigate how code search differs from non-code related search with \GPSE in two dimensions: (i) query characteristics (RQ1), and (ii) users' effort (RQ2 and RQ3). To this end, we explore the following research questions:

%\RQA{1}{ How effectively can a \cscore capture the code intent of the query?} 
%\noindent In this RQ, we discuss our \textit{codeness} threshold procedure and evaluate our intent based classifier's performance on manually annotated queries.

\RQ{1}{ How do query characteristics differ for code and non-code queries?} 
\noindent 
%Users convey their informational need to the search engine by issuing a query. To give satisfactory results, it is important for a search engine to understand query characteristics. 
To explore this RQ, we analyze how linguistically the two queries are different. In particular, we check whether query length varies for code and non-code search. We also study the vocabulary sizes and vocabulary choices between the two. 
% Usually, a source code query is intended for a particular programming language (\ie Java, C\#). %, which is an important factor to know to generate effective search results. 
% We also analyze how often users mention the language by name in code related queries.
% ~\md{This does not make much sense. It would be better to give statistics found by Jed. Also, what are the top key words for code vs non-code queries?}

\begin{comment}
\begin{enumerate}
    \item How do code-related queries differ in length from non-code-related queries?
    %\item How common are the most frequent code-related queries compared to the most frequent non-code-related queries? (i.e. how diverse are both types of queries?
    %\item How often do developers search with the direct API name and other code related tokens in the query?
    \item How often do developers mention a programming language in a code query?
\end{enumerate}

\end{comment}

\RQ{2}{ How do search behaviors vary for code and non-code related queries?} 
\noindent 
%Understanding users' search behavior is vital for any search engine's success. 
In this RQ, we explore different search behaviors of users, including how much time they spend on search results, how many websites they visit, and how often they modify their search queries. We also analyze how this behavior varies for code and non-code related searches.

\begin{comment}
\begin{enumerate}
    \item How long do users spend searching for code-related issues compared to non-code-related issues?
    %\item How long do users spend on each website while searching for code-related issues compared to non-code-related issues?
    \item How many websites do people traverse when searching for code-related issues compared to non-code-related issues?
    \item How often do people have to modify their search when searching for code-related issues compared to non-code-related issues?
\end{enumerate}

\end{comment}

\RQ{3}{ How do task sessions vary for code and non-code related search tasks?} 
\noindent 
Often, several queries can be related to same web search task. To explore this RQ, we identify sequences of related queries as \emph{task sessions} (Section~\ref{sec:data-collection}). Next, we analyze how many queries, how much time, and how many website visits users require to complete a task. We also analyze how these task level interactions differ for code related search compared to non-code.

\begin{comment}
    \begin{enumerate}
        \item How many query users need to complete a search task?
        \item How much time do users need to complete a search task?
        \item How many different website visits are required to complete a search task?
        %\item How do users modify queries within a search task?
    \end{enumerate}
\end{comment}

%These observations might be useful to pinpoint the weaknesses of existing code search techniques and thus help to design improved code search tools. 

%!Tex root=main.tex
\section{Code Intent Analysis}
\label{codeness-model}
%After collecting query data from the search log, we analyze the code intent of each query. 

%If a query token has a code intent, we consider that as a code token; for example, "javascript", "C\#", "json", "visual-studio", etc. 
We assume that if a query contains more code related tokens (\eg "javascript", "C\#", "json", "visual-studio"), it indicates more code intent. To automatically estimate such intent, we build an analysis technique that assigns a code intent score to each query. We call this score as \cscore. %The model first identifies code intent of tokens and then leverages these tokens to assign a \cscore of a query.
In this section, we discuss our analysis technique in details. 

\subsection{Code Intent of Tokens}
To construct the model, we first collect a list of code related token set ($S$). We leverage StackOverflow (SO)~\cite{stackoverflow} (May 2017 data dump) which is an online Q/A forum where developers often discuss their programming related issues. A post in SO can be associated with tag(s), which are given manually. 
%These tags can be thought of as a comprehensive list of programming problem-related tokens or code tokens. We consider all the tags that occur in at least one SO post as our code token set $S$ (SO data dump May 2017). 
However, not all the tags are equally strong indicators of code intent. We deal with such scenarios as follows: 

Firstly, we filter out ambiguous tags from our token set. 
SO tags often co-occur with other tags and thus there might exist some tags which always co-occur and never occur alone in any post. These tags might not be an indication of code token. For example, "unbox" tag never occurs alone but occurs with "haskell" tag~\cite{so-post:haskell-unbox} which is a code token. Such tag (\ie "unbox") might not be an indicator of code intent. To remove such unwanted noise, we filter out all the tags which never occur alone in any post. Additionally, we remove all the post with multiple tags. Thus the frequency of a tag is the count of its single occurred posts only. This process reduced the number of selected tags drastically from 
$46.3K$ %$46278$ 
to
$19.8K$ %$19752$.

Secondly, we assign a \cscore for each tag in our filtered code token set ($S$). We assume that the popularity of a tag on SO is the indicator of its code intent. Higher frequency (\ie popularity) indicates strong code intent. However, the raw frequency might lead to incorrect code intent estimation.  For example, in Table~\ref{tab:tag-codeness}, the frequency (\ie count) difference between "android" and "java" shows "android" carries much higher code intent than "java" which is not completely accurate estimation. To mitigate such frequency difference bias, we use sub-linear scaling. If a tag $x$ occurs $n$ times then the \cscore, $f(x)$, of that token is given by equation~\ref{eqn:token-score}, 
\begin{equation} \label{eqn:token-score}
f(x) =
\begin{cases}
1+log2(n), & \text{if }x\in S \\
0, & \text{if }x \notin S
\end{cases}
\end{equation}

\noindent where $S$ is the code token set. 
Note that, if a token is not in the code token set ($S$) its \cscore is $0$ and that token is considered as a non-code token. $n$ is the frequency of token x across all Stack Overflow posts.

Now, considering previous "android" vs "java" example, we see the \cscore are $17.55$ and $17.13$ (in Table~\ref{tab:tag-codeness}) which shows both tag are of similar code strength. In contrast, in Table~\ref{tab:tag-codeness} \cscore of "lucene" is $10.18$, which indicates its code intent is less compared to "android" or "java". Some code tokens in different count ranges, and their \cscore can be found in Table~\ref{tab:tag-codeness}.

%For instance, the frequency\footnote{as of May 2017} of "C\#" and "solr " are $48898$ and $2236$ respectively and their \textit{codeness score} are $16.58$ and $11.13$ respectively. 
%The frequency of "C\#" and "solr " are $48898$ and $2236$ respectively which implies, code intent of "C\#" is stronger than "solr".  is large and "C\#" is around $21$ times popular than

% Table generated by Excel2LaTeX from sheet 'Sheet2'
\begin{table}[t]
  \scriptsize
  \centering
  \caption{Sample code tokens' count (single occurrence) and their \cscore}
    \begin{tabular}{lll|lll}
    \toprule
    Tags  & count & cscore & Tags  & count & cscore \\
    \midrule
    android & 96210 & 17.55 & css3  & 982   & 10.94 \\
    java  & 71869 & 17.13 & applescript & 956   & 10.9 \\
    php   & 71390 & 17.12 & lucene & 579   & 10.18 \\
    javascript & 70248 & 17.1  & coffeescript & 579   & 10.18 \\
    python & 53993 & 16.72 & firefox-addon & 268   & 9.07 \\
    jquery & 52705 & 16.69 & livecode & 268   & 9.07 \\
    c\#   & 48898 & 16.58 & jasmine & 86    & 7.43 \\
    mysql & 41684 & 16.35 & codeigniter-3 & 86    & 7.43 \\
    c++   & 41283 & 16.33 & miniprofiler & 4     & 3 \\
    r     & 30176 & 15.88 & idocscript & 1     & 1 \\
    \bottomrule
    \end{tabular}%
  \label{tab:tag-codeness}%
\end{table}%
\subsection{Code Intent of Queries}
We leverage the token level codeness score to compute the \cscore of the query. The \cscore (\textit{cscore}) of a query is calculated by summing up the code score of its tokens as in equation \ref{eqn:query-score}

\begin{equation} \label{eqn:query-score}
cscore(Q) = \sum_{i=1}^{m} f(x_i)
\end{equation}
where $x_i$ is the $i^{th}$ token of the query $Q$ of length $m$ and $f(x_i)$ is the \cscore of token $x_i$ as in equation ~\ref{eqn:token-score}.

If the \cscore of a query is high it is considered to have a high code intent. In this way, the model assigns a code intent to each query. Some sample queries with their \cscore is shown in  Table~\ref{tab:sample-query-code-score}.

% Table generated by Excel2LaTeX from sheet 'codeness-example'
\begin{table}[t]
  \scriptsize
  \centering
  \caption{Sample query and their codeness score assigned by our model}
    \begin{tabular}{rp{11.07em}cp{4.215em}}
    \toprule
          & Query & {Code Score} & Type \\
    \midrule
    1     & javascript mp3 play time & 40.71 & Code \\
    \midrule
    2     & javascript get track length from meta data & 48.57 & Code \\
    \midrule
    3     & how to perform xml serialization for parameterless constructor in c\# & 67.33 & Code \\
    \midrule
    4     & elasticsearch.net \& nest installed post nuget source control stop notification & 49.36 & Code \\
    \midrule
    5     & acer e700 review & 7.07  & Noncode \\
    \midrule
    6     & houston luxury suv rental & 0.00  & Noncode \\
    \midrule
    7     & messi curly goal & 2.58  & Noncode \\
    \bottomrule
    \end{tabular}%
  \label{tab:sample-query-code-score}%
\end{table}%

%!Tex root=main.tex

\section{Methodology}
\label{sec:method}

In this section, we start with explaining our data collection and query extraction approach in detail. Then we present both manual and automatic query annotation process. After that, we describe search tasks extraction and classification method.
%!Tex root=main.tex
\subsection{Study Subject}
\label{sec:data-collection}

%\subsection{Dataset Details}
%We first collect the search log data, then identify queries and corresponding clicked URLs from the activity log data.

%\textbf{Search Log}.

Our search log data was collected from developers who installed a proprietary Google Chrome Web Tracking plugin~\cite{codealike}.
%\footnote{Plugin name and reference blinded for review} %The plugin is a service that tracks developers' activity while they code. 
The plugin tracks all the web browsing activities which are processed and analyzed to understand how developer work and learn. %~\codealike{verify and add content if required}
Thus in our dataset most of the users are developers either acting as team leaders or performing technical tasks and the activity includes search query and clicked web page visit information. 

The data collection period spanned $14$ months starting from December 2014 to January 2016.  
There are a total of $149,610$ queries of $310$ users (See Table~\ref{tab:dataset-edit-query-user}). %~\todo{check whether this info violates double-blind review requirement}

%including software development related queries (code) and general search queries (non-code). 
%This setting restricted users to query only on Google Search Engine. 
%\codealike{verify and add any relevant information about the dataset here}

% \subsection{Terminology}
% \label{query-extraction}
The dataset contains information about \emph{activity sessions} for each user, which represent a user's active development time~\cite{corley2015web}. Each user can have many activity sessions.  Each activity session contains events in the web browser, such as search queries and results clicked, as well as non-browser events, such as IDE interactions. 
For the browser events,  the logs provide information on the clicked results, specifically, the URL and page title. All events have a start time and an end time. 

% We first identify the search queries and corresponding clicked URLs from the search results. 
These activity sessions provide a useful boundary of continuous activity for a user, but finer granularity is needed as the logs contain information about browser interactions as well as non-browser interactions. Further, we want a notion of related web activities, since users often initiate consecutive yet unrelated search queries. After identifying consecutive related queries, 
we split the activity sessions into \emph{task sessions}. 
% Thus, these task sessions represent sequences of related search queries and web browser interactions. 
This is accomplished by first identifying \emph{edited queries}.  

\noindent
\textbf{Identifying Edited Query}:
Users often modify their search query to give more specific information to the search engine.
These query reformulations can expand the query by adding more terms or reduce the query by removing terms. 
% their search query to better express their search intent to the search engine.  
If a query contains \textit{at least one common term} with its previous query, and the queries come from the same activity session, we consider both queries as edited queries.
%, which covers both expansion and reduction reformulations~\cite{haiduc2013automatic}. \\

% Nevertheless, a user might search for a query and not click any results. In that case, we consider only search page exploration time. Thus, to get total query search time, we consider the total time which starts from issuing a query and ends when next query is issued within an activity session~\md{activity session is never defined earlier}. Details of query extraction can be found in Section~\ref{query-extraction}.\\

% \noindent
% \textbf{Code Query}:

\noindent
\textbf{Composing Task Sessions}:
Task sessions capture continuous, related web browser interactions. 
We consider all browser events after one search query and before the next query as the {\em result exploration} activity for the former query; the web URLs of those activities are considered \textit{clicked URL}. 

To identify task sessions, first we explore all continuous sequences of edited queries and their associated results exploration activities. Each sequence of edited queries represents a task session. 
The remaining queries are all non-edited. Each non-edited query, along with its results exploration activities, forms its own task session. 

% we collect continuous web browser interactions (queries, clicked URLs) such that the start time of the next activity is the same as the end time of the previous activity. \\
% Users often turn to a search engine to know about a particular issue, such as: a fact (\eg what is the current temperature, what is the latest version of Android OS), knowledge seeking for a particular topic, or problem-solving. Users often interact with the search engine to complete such tasks. The interaction might be observing results from the first query and gathering some knowledge from the results, and then submitting another query to get more information. The next query might be an edited query from the first query. Thus, users gradually get to know about the complete solution through multiple interactions with the search engine~\cite{martie2017understanding}. After a certain interaction, users may end up with getting complete results or remain unsatisfied and give up on searching for that topic. We consider these activities as a task and this time span as a task session.
% \md{it is still not clear how a task session is identified. Same user, obviously, but beyond that, how did you delineate between sessions? Was it a measure on idle time between interactions? or something else?}

\noindent
\textbf{Computing Search Query Time}:
Users spend time on the search page and on the web pages they click. The time between when a query is issued and when the next query is issued, or the activity session ends (whichever comes first), is referred to as the \emph{query search time}. In the event that a user does not click any results, the \emph{query search time} is computed as the time spent on the result page. 

% \subsubsection{Queries}
% The user expresses their information need in terms of a query, and the search engine returns a list of the documents which are relevant to that query. Typically, each returned document consists of web page title, URL, metadata (\ie date), and a short summary of the web page. 
% This summary usually explains why that document is relevant to the query (\ie highlight query keywords).\\ 

% \noindent
% \textbf{Query Search Time}:

%Similarly, we consider the clicked URLs as \textit{websites visit} for each query.
%Note that, we also consider task session, which comprises of multiple queries (in RQ\ref{rq4}).

%!Tex root=main.tex
%\subsection{Manual Query Annotation and Classifier Evaluation}
\subsection{Query Classification}
\label{query-classification}
A query which is intended to solve any software development related issue is considered as a \textit{code} query. For example, \textit{reference code example} (\eg "write in file java", and "how to get all textbox names inside table layout panel c\#"), debugging (\eg "asp.net mvc error page"), API usage, technical knowledge (\eg "npm update all dependencies", "git bash mingw", and "qualities of good programmer") and other development related tasks are considered as code related query.
A query which is not intended to solve any software development or programming task is considered as a \textit{non-code} or general query. For example, "make your own comics", "review Galaxy Note Edge", and "d5300 amazon" are considered as non-code.
To set the threshold and evaluate our classifier, we manually annotate queries to code and non-code.

\textbf{Manual Query Annotation}. 
 From our dataset, we randomly sample $380$ queries across users. Two researchers separately annotated those queries and resolved the disagreement by discussion. 
%We find a mutual agreement of $93\%$ between annotators. 
We measure Cohen's kappa coefficient ($\kappa$)~\cite{cohen-kappa} to find inter-annotator agreement, where a $\kappa$ value of $1$ indicates a complete agreement and a value of $0$ indicates a complete disagreement. In our annotation we find a $\kappa$ value of $0.85$. \\%  ~\todo{use the Cohen's Kappa coefficient here too}

%We evaluate our classifier on the manually annotated query and use following metrics to measure classifier's performance.
%\noindent
\noindent \textbf{Evaluation Metrics}. 
We use following metrics to evaluate our classifier: 

\noindent
\textit{Precision (P) - } is the fraction of correct prediction of total query. Thus, $P = \frac{r}{d}$, where $r$ is the number of correct prediction and $d$ is the total query.

\noindent
\textit{Recall (R) - } is the fraction of correct prediction of the total ground truth. If $t$ be the total ground truth, the recall is $R = \frac{r}{t}$. 

\noindent
\textit{F$_1$ Score (F-1) - } is a single combined metric that trades off precision vs.~recall by computing the harmonic mean of the two: \label{f-1-score-eq} 
$F_1 = 2 * \frac{{precision}*{recall}}{{precision}+{recall}}$.

\begin{figure}[!htpb]
\centering
\scriptsize
 \includegraphics[width=0.9\columnwidth]{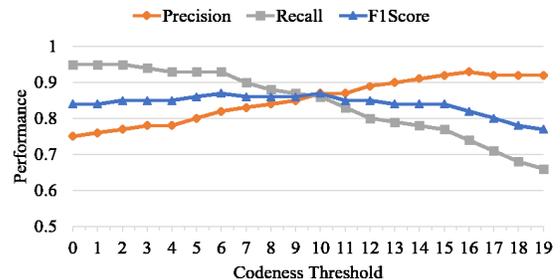}
 \caption{\textbf{\small Classifier Evaluation}}
 \label{fig:classifier-eval}
 \vspace{-0.1cm}
\end{figure}

\noindent \textbf{Accuracy Evaluation}.
% \textbf{How effectively can a \cscore capture the code intent of the query?}
Figure \ref{fig:classifier-eval} shows \textit{precision}, \textit{recall}, and \textit{F1-score} with respect to \textbf{code query} in different \cscore thresholds. As the threshold increases, \textit{precision} also increases. % thus the classifier become more conservative. 
In contrast, \textit{recall} decreases with the increase in threshold. %, thus incorrectly classifying more code queries. 
However, \textit{F1 Score} remains in between \textit{precision} and \textit{recall} in different thresholds. For a better comparison, it is important to maintain a balance between code and non-code query classification. Thus, we choose the threshold = $10$ where the classifier achieve a better trade-off of $Precision = 87\%$, $Recall = 86\%$, and $F1 Score = 87\%$.

These results indicate that our model assigns \cscore which is effective in detecting query intent.
In addition to separating code from non-code, the model also identifies the code specificity of a code query. 
%The \cscore is a continuous value that can identify the strength of the query's relationship to code.
A larger score indicates a strong code intent. Thus, we can further separate code related queries into different clusters based on different ranges of \cscore.  

%\RS{1}{Codeness score effectively captures code intent and identifies code query with $Precision = 87\%$, $Recall = 86\%$, and $F1 Score = 87\%$.}

%\noindent
%\textbf{Automatic Query Classification.} 
For our analysis, we need to classify all queries to code or non-code. 
%Unfortunately, our dataset comes with a search log without such query annotations.
% We leverage \cmodel (details in Section~\ref{codeness-model}) to build a classifier which automatically identify code related query from non-code related query. For a query, higher \cscore refers to higher code intent. On the other hand, smaller \cscore indicates less code intent and score smaller than a certain threshold considered as the non-code related query.
Empirically, we set \cscore threshold to 10, which gives us a better trade-off for precision and recall. Details dataset statistics after the query classification can be found in Table~\ref{tab:dataset-edit-query-user}.

\begin{comment}
Some examples from our dataset as follows\todo{can be included in dataset description/ classification model discussion also, the codenesss score can be added}: 
Code-related queries, codeness score in "()": 

\begin{itemize}
    \item \textit{javascript get track length from meta data ($48.57$)}
    \item \textit{how to perform xml serialization for parameterless constructor in c\#}
    \item \textit{elasticsearch.net \& nest installed post nuget source control stop notification}
    \item javascript mp3 play time (40.7081804594)
javascript get track length from meta data (48.5740057299)
how to perform xml serialization for parameterless constructor in c# (67.3328320023)
elasticsearch.net \& nest installed post nuget source control stop notification (49.3627378535)
acer e700 review (7.06608919046)
houston luxury suv rental (0.0)
messi curly goal (2.58496250072)
\end{itemize}

Non-code-related queries
\begin{itemize}
    \item \textit{acer e700 review}
    \item \textit{pepsi flavored cheetos youtube}
    \item \textit{j cole bachelor degree}
\end{itemize}
\end{comment}

\begin{comment}
% Table generated by Excel2LaTeX from sheet 'user-query-stat-edit'
\begin{table}[htbp]
  \centering
  \caption{Add caption}
    \begin{tabular}{ll}
    \toprule
    Total Query & 149610 \\
    Total Users & 310 \\
    Query/User & 482.61 \\
    \bottomrule
    \end{tabular}%
  \label{tab:addlabel}%
\end{table}%

\end{comment}

% Table generated by Excel2LaTeX from sheet 'data-stat'
\begin{table}[t]
\scriptsize
  \centering
  \caption{Dataset Statistics (Codeness Score Threshold $= 10$)}
    %\begin{tabular}{rcc|cc|cc|rrrrr}
        \begin{tabular}{ccc|cc|rrrrr}
    \toprule
          &       &       &       & Query/ & \multicolumn{5}{c}{User-Query Stats} \\
    Query & \#    & \%    & User  & User  & \multicolumn{1}{c}{Min} & \multicolumn{1}{c}{Q1} & \multicolumn{1}{c}{Q2} & \multicolumn{1}{c}{Q3} & \multicolumn{1}{c}{Max} \\
    \midrule
    Code  & 88577 & 59.21 & 300   & 295.26 & 1     & 22    & 136.5 & 387.25 & 2593 \\
    Noncode & 61033 & 40.79 & 296   & 206.19 & 1     & 15    & 85.5  & 294   & 1642 \\
    \midrule
    All   & 149610 & 100   & 310   & 482.61 & 1     & 28.75 & 207   & 676   & 3632 \\
    \bottomrule
    \end{tabular}%
  \label{tab:dataset-edit-query-user}%
\end{table}%

%\subsection{Extract Query Search Time and Websites Visit}The user expresses their information need in terms of a query, and the search engine returns a list of the documents which it thinks are relevant to that query. Typically, each returned document consists of web page title, URL, metadata (\ie date), and a short summary of the web page. This summary usually explains why that document is relevant to the query (\ie highlight query keywords). In a sequential search setting (\ie Google), it has been assumed that users search from top to bottom~\todo{ref, if any} and click results which they think are relevant. The user spends some time exploring content on the clicked web page and returns back to the search engine to gather more information about the query. In a nutshell, users spend time on the search page and on the clicked web pages. We consider this total time as the time for this query. Note that a user might search for a query and not click any results. In that case, we consider only search page exploration time. We apply a similar procedure to collect the clicked URLs for each query.   

\subsection{Extract and Classify Search Task}

We analyze how user interaction varies for code and non-code related search tasks. First, we extract task information from the search log data. 
We define a search task as a set of the consecutive edited queries. 
We start with a query and consider all the subsequent queries which were edited from the previous query and stop when encountering a totally new query. 
Thus we extract all such tasks for all users. Applying this process, we extract a total of 108,313 tasks. 
Secondly, to analyze code and non-code task properties, we compute the \cscore for each task. We assign a representative query whose \cscore is maximum among all other queries in a task. We consider this maximum \cscore as the code intent for that task. Thus we assign a \cscore for all the tasks. Similar to query classification, we consider a task with a \cscore greater than a particular threshold ($10$ in our experiment) as code related task and non-code otherwise.
Note that, in a task, a user might start with a lower code intent query and can add code token(s) gradually to increase the code intent of the whole task. Sample task session from our dataset can be found in Table~\ref{tab:task-sample}.

% Table generated by Excel2LaTeX from sheet 'task-sample'
\begin{table*}[t]
 \scriptsize
  \centering
  \caption{Sample Task Sessions (from Dataset)}
    \begin{tabular}{lclll}
    Task & Edit Seq. & Query & Added Terms & Deleted Terms \\
    \midrule
        & 1     & how to get mp3 playtime in c\# from stream & \multicolumn{1}{l}{} & \multicolumn{1}{l}{} \\
          & 2     & javascript mp3 play time & javascript, play, time  & how, to, c\#, from, stream, playtime \\
  Code    & 3     & how to get mp3 play time length & how, to , get, length & javascript \\
          & 4     & javascript function to get mp3 play length & javascript, function & how , time \\
          & 5     & javascript read mp3 metadata & read, metadata & function, to, get, play, length \\
    \midrule
    Noncode & 1     & enterprise luxury suv & \multicolumn{1}{l}{} & \multicolumn{1}{l}{} \\
          & 2     & luxury suv rentals houston & rentals, houston & enterprise \\
    \bottomrule
    \end{tabular}%
  \label{tab:task-sample}%
\end{table*}%

\subsection{Codeness Difference Calculation}

Search engines (\ie Google) often suggest query edits with the search results. This helps users to come up with their desired query for their informational need. To this end, we analyze what happens  \wrt \cscore when users edit a code related query.  
%Note that the \cscore is linear with the number of code tokens in the query and the\cscore of an edited query increases in the following cases: i) insert code-token(s), and ii) replace non-code-token(s) with code-token(s) (delete+insert). On the other hand, \cscore decreases in the following cases: i) delete a code-token(s), and ii) replace code-token(s) with non-code-token(s) (delete+insert). Lastly, if a user inserts or deletes non-code-token(s) only then does \cscore remain the same for the edited query.
If a query $q$ is reformulated to $q_r$ then we calculate their \cscore difference, $\Delta Codeness$, as in Equation~\ref{eqn:delta-codeness}.
\begin{equation} \label{eqn:delta-codeness}
\Delta Codeness = Codeness(q_r) - Codeness(q)
\end{equation}

Here, a positive value of $\Delta Codeness$ indicates an increase, a negative value indicates a decrease and zero ($0$) value indicates no-change in code intent after reformulation. We compute the $\Delta Codeness$ for all the edited code related queries in three different settings: edited 1) only by adding term, 2) only by deleting term, and 3) overall, adding or/and deleting term.

%\input{classification.tex}
%!Tex root=main.tex

\section{Results}
\label{sec:exp}
In this section we discuss our experiments and results analysis of RQs.

%\subsection{Results and Analysis}
%\input{rq1-old.tex}

%!Tex root=main.tex

\RQA{1}{ How do query characteristics differ for code and non-code queries?} 
\label{rq1}

To analyze query characteristics we filter out the duplicates to mitigate unwanted bias from query duplication. In this search log, we found $20.36\%$ duplicate queries with duplication for both code ($20\%$) and non-code ($21\%$) queries.

\textbf{1) How do code-related queries differ in length from non-code-related queries?}

\begin{figure}[!htpb]
\centering
\scriptsize
 \includegraphics[width=0.9\columnwidth]{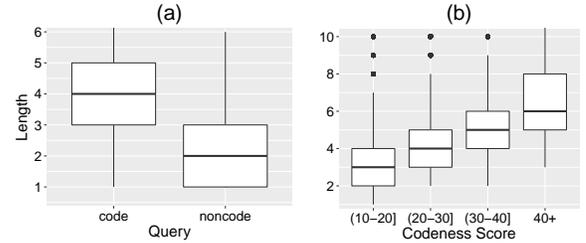}
 \caption{\textbf{\small Query Length (\# of words)}}
 \label{fig:qlength-combine}
 %\vspace{-0.1cm}
\end{figure}

We begin with exploring the query length. Figure~\ref{fig:qlength-combine}(a) shows that code related query length (\ie number of tokens or words in a query) is often higher than the non-code, with statistical significance (confirmed by Wilcox statistical significance test with \textit{medium} Cohen'D effect size). The average length ($4.7$) of the code related queries is higher than non-code ($2.3$). This implies that users tend to use more words to express a code related issue which is almost twice that needed to for a general non-code issue. 
%As the query gets longer its intent becomes more specific. 

%This result implies that code related queries are more specific than non-code related queries and users often use a longer query for code related search compared to general-purpose search. 

%A longer query can be thought of as an indication of specificity of the query intent~\todo{ref}. The result indicates that users often search with more specific intent for code related search than general purpose non-code search.

To dig into this further, we analyze how the query length varies with the increase of the code intent of the query (\ie \cscore). Figure ~\ref{fig:qlength-combine}(b) shows the comparison in query length in different \cscore ranges. We see that often, queries with higher \cscore are longer in length. Note that by definition \cscore increases with the increase of code related tokens in a query. However, adding a non-code token does not an increase in \cscore. Thus, sharp increase of query length with \cscore in Figure ~\ref{fig:qlength-combine}(b) confirms that code related query are indeed more verbose. 

%These examples clearly depict the longer queries that are required for code-related searching compared to non-code-related searching.
%\textbf{Result Summary- RQ-1-1:} Users use more words when searching with code-related queries than they do with non-code-related queries.
 
%\item How common are the most frequent code-related queries compared to the most frequent non-code-related queries? (i.e. how diverse are both types of queries?
%\item How often do developers search with the direct API name and other code related tokens in the query?

\begin{figure}[!htpb]
\centering
 \includegraphics[width=0.9\columnwidth]{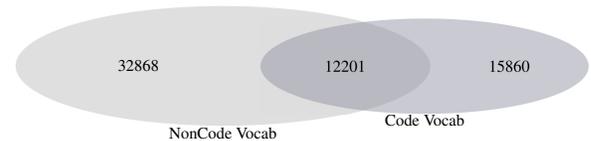}
 \caption{\textbf{\small Vocabulary words statistics for code and non-code query}}
 \label{fig:vocab-stat}
 \vspace{-0.1cm}
\end{figure}

% vocab results
\textbf{2) How do vocabulary varies for code and non-code query?}
%We analyze how vocabulary varies for code and non-code query. 
For this analysis, we remove English stop-words (adopted from ~\cite{english-stopword-link}) from the queries . We find that vocabulary of code is $28K$ which is much smaller than non-code query $45K$ though, in our annotated data, the number of code query (\ie $59.21\%$) is higher than non-code (\ie $40.79\%$). 
However, we observe that $43.48\%$ of code vocabulary are common with non-code which is depicted in Figure~\ref{fig:vocab-stat}. In Table~\ref{tab:vocab-top} we list top frequently occur code, non-code, and common tokens with their frequency.

% Table generated by Excel2LaTeX from sheet 'vocab-compare'
\begin{table}[t]
\scriptsize
  \centering
  \caption{Top query words and frequency for Code, Noncode and common words between them (w/o English Stopword)}
    \begin{tabular}{lr|lr|lr}
    \toprule
    \multicolumn{2}{c}{Code} & \multicolumn{2}{c}{Common} & \multicolumn{2}{c}{NonCode} \\
    Token & \multicolumn{1}{l}{Freq} & Token & \multicolumn{1}{l}{Total Freq} & Token & \multicolumn{1}{l}{Freq} \\
    \midrule
    c\#   & 6165  & string & 1179  & 2015  & 262 \\
    sql   & 2604  & add   & 982   & de    & 204 \\
    windows & 2587  & type  & 950   & define & 153 \\
    javascript & 1966  & error & 855   & meme  & 108 \\
    server & 1936  & create & 847   & uk    & 104 \\
    jquery & 1713  & change & 844   & dell  & 99 \\
    studio & 1696  & list  & 826   & la    & 94 \\
    visual & 1639  & set   & 809   & day   & 83 \\
    file  & 1443  & object & 782   & price & 82 \\
    string & 1160  & array & 741   & world & 79 \\
    mvc   & 1144  & table & 695   & south & 79 \\
    web   & 1038  & 2015  & 687   & movie & 79 \\
    code  & 979   & date  & 656   & lyrics & 76 \\
    add   & 946   & find  & 645   & weather & 75 \\
    type  & 929   & time  & 633   & top   & 73 \\
    asp.net & 915   & check & 627   & road  & 73 \\
    \bottomrule
    \end{tabular}%
  \label{tab:vocab-top}%
\end{table}%

%\textbf{3) How often do developers mention a programming language in a code query?}

Code related queries are often intended for a particular programming language. To reduce the search space for relevant documents, it is important to understand what programming language the user intended to use.
To explore this, we analyze how frequently user explicitly mentions the language by name in the query. We use a list of $100$ popular programming languages~\cite{pl-popular-100} and search for these keywords in code related queries. We find that users mention language name in the code query $20\%$ of the time and skip $80\%$ of the time. 
This indicates that most of the time developers do not mention language explicitly thus it is up to the search engine to guess which programming language the developers intended.
The top mentioned programming languages with their query frequency is shown in Table \ref{tab:pl-mention-freq}.

% Table generated by Excel2LaTeX from sheet 'rq_9'
\begin{table}[t]
\scriptsize
  \centering
  \caption{Most Frequently Mentioned PLs. $20.10\%$ of code queries mention at least one PL name}
\begin{tabular}{clc|clc}
    \toprule
    Top   & Language & Freq. & Top   & Language & Freq. \\
    \midrule
    1     & c\#   & 6274  & 11    & python & 192 \\
    2     & sql   & 2586  & 12    & c     & 145 \\
    3     & javascript & 1970  & 13    & bash  & 131 \\
    4     & .net  & 683   & 14    & go    & 130 \\
    5     & php   & 469   & 15    & ruby  & 102 \\
    6     & powershell & 388   & 16    & crystal & 94 \\
    7     & assembly & 267   & 17    & r     & 87 \\
    8     & c++   & 255   & 18    & logo  & 58 \\
    9     & java  & 250   & 19    & s     & 56 \\
    10    & icon  & 203   & 20    & f\#   & 53 \\
    \bottomrule
    \end{tabular}%
  \label{tab:pl-mention-freq}%
\end{table}%

%\RS{1}{Users 1) search for code by issuing longer queries whose length is, on average, nearly double that of the length of a general search query, and 2) skip programming language's name in around $80\%$ of their queries.}

% \RS{1}{
% 1) Code query is often longer than non-code and, on average, nearly double the length of a general search query,
% 2) Code vocabulary is smaller than non-code and less than half of the code query token are common with non-code, 
% and 3) Almost $20\%$ of code queries contain  programming language name.}

\RS{1}{
 Code queries are linguistically different than non-code queries\textemdash they are longer and contain less vocabulary.
 %linguistic often longer than non-code and, on average, nearly double the length of a general search query,2) Code vocabulary is smaller than non-code and less than half of the code query token are common with non-code
 }
%!Tex root=main.tex
\RQA{2}{ How do search behaviors vary for code and non-code related queries?} 
\label{rq2}

We investigate this question in three dimensions: (i) how much time users have to spend per query, (ii) how many websites they visit per query, and (iii) how many times users have to edit a query to retrieve the intended document. We will discuss them one by one.

\textbf{1) How long do users spend searching for code-related issues compared to non-code-related issues?}

\begin{figure}[!htpb]
\centering
\scriptsize
 \includegraphics[width=0.9\columnwidth]{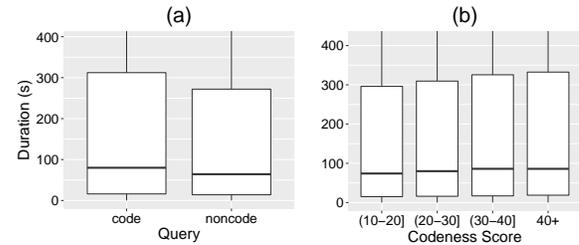}
 \caption{\textbf{\small Query Browsing Duration}}
 \label{fig:query-duration}
 \vspace{-0.1cm}
\end{figure}

We compare time duration for code and non-code query in Figure~\ref{fig:query-duration} (a). We see that in code related query users take slightly more time compared to non-code query search with the median time of $1$min $20$ sec and $1$min $4$ sec for code and for non-code queries respectively. Although this difference is statistically significant (Wilcox's Test), CohenD's effect size is negligible.

We further check whether time duration varies with \cscore. Figure~\ref{fig:query-duration} (b) shows that as the \cscore of queries increases users tend to spend slightly more time on searching (with negligible effect size). 
Thus, in reality, we do not see any major difference code and non-code queries \wrt the time users spend interacting with the search engine.

%\textit{Experiment Setup:}
%The data set for this experiment documented the amount of time a user spent on each page. Using this information, we found the length of a user's browsing session for a certain query by adding up the times that the user spent on each page searching for that query. We then generated box plots for the average session for both a code-related query and a non-code-related query.

%Similar browsing session for code and noncode query indicates, users spend similar amount of time to find relevant information from a search results. This characteristics might impose additional challenges for code related search as for even difficult query, user might expect equally good results as other general purpose query.

%However, in RQ~\ref{rq2}, we observe that higher code intent queries are often longer in length, and thus are more specific. Results in here in Figure~\ref{fig:query-duration} (b) indicates that users spend more time for higher code intent queries. Though the higher code intent queries are specific, a search engine's results might not be effective and thus the user ends up spending more time on those queries.

%Average duration per webvisit is $23.41$ second for code query and $19.35$ second for non-code query.

%\item How long do users spend on each website while searching for code-related issues compared to non-code-related issues?

\textbf{ 2) How many websites do people traverse when searching for code related issues compared to non-code related issues?}

\begin{figure}[!htpb]
\centering
\scriptsize
 \includegraphics[width=0.9\columnwidth]{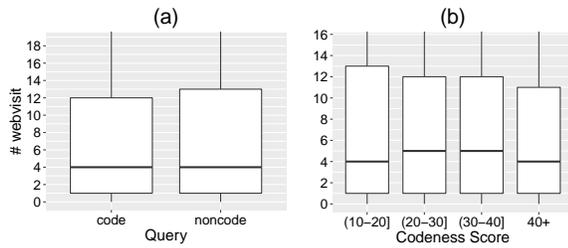}
 \caption{\textbf{\small Web Visit Analysis for Query}}
 \label{fig:webvisit-combine}
 \vspace{-0.1cm}
\end{figure}

From Figure \ref{fig:webvisit-combine} (a), we see that there is no significant difference (confirmed by Wilcox statistical significance test and Cohen'D effect size) between code and non-code with a median of $4$ for both query types. The average number of clicks per query is $11.4$ for code and $12.8$ for non-code. This result indicates no matter what types of problem users search for, they often visit a similar number of websites. One possible explanation is - after a certain number of clicks, users stop exploring results no matter whether the returned results are satisfactory or not.
This hypothesis can be explained further by the results in Figure~\ref{fig:webvisit-combine} (b). We see that there is no visible trend of the number of website visits in the different range of \cscore. 
%We can conclude that how many clicks per query is a common behavior of users. 
%\todo{ref other related survey results - IR if any} 
We can conclude that users visit a similar number of website in general for all queries. This behavior can be considered as a common search behavior. 

Furthermore, we observe that $22\%$ of non-code queries require no website visits, which is higher when compared to $17.9\%$ for code related search. This indicates that for non-code general search, users get relevant results from the search results page only (\ie fact searching, summarized results from Google) or quickly realize whether the resulting information is relevant at all. On the other hand, for code related search users need to click and see the content (most of the cases) of the website to judge whether the page contains relevant information or not.

\textbf{ 3) How often do people have to modify their search when searching for code related issues compared to non-code related issues?}

Depending on the the context, the user might add/insert terms to the query, delete terms, or both in order to reformulate the query. Such queries are called edited queries. The user keeps editing a query repeatedly until they are satisfied with the returned results. Thus, an ideal search engine would return the exact satisfactory documents when user issues a query for the first time. The more a query is edited, the more effort is needed from the user to find out relevant documents. To this end, we observe how users modify queries during searching.

We observe that in total $27.6\%$ of queries are edited queries. In particular, $34.9\%$ of the code queries are edited queries, which is significantly higher than $17.01\%$ of non-code edited queries. % (in Table~\ref{tab:dataset-edit-query-user}).
This result indicates that search engine (\ie Google) found it twice as difficult to understand code search compared to non-code search.

\begin{figure}[!htpb]
\centering
\scriptsize
 \includegraphics[width=0.9\columnwidth]{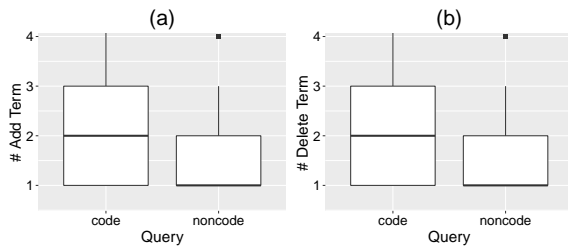}
 \caption{\textbf{\small Add and Delete Term in Query Statistics for Code vs Non-code}}
 \label{fig:edit-add-delte-term}
 \vspace{-0.1cm}
\end{figure}

% Table generated by Excel2LaTeX from sheet 'query-terms'
\begin{table}[t]
\scriptsize
  \centering
  \caption{Term Statistics for Edited Code Query }
    \begin{tabular}{ll|ll}
    \toprule
    \multicolumn{2}{c}{Top Added Terms} & \multicolumn{2}{c}{Top Deleted Terms} \\
    Unigram & Bigram & Unigram & Bigram \\
    \midrule
    to    & visual studio & to    & visual studio \\
    c\#   & c\# winforms & in    & when only \\
    in    & framework entity & for   & success is \\
    the   & not working & and   & status code \\
    of    & windows 8 & with  & returned success \\
    windows & how to & c\#   & phone css \\
    sql   & model object & on    & parameters in \\
    a     & unit test & object & not working \\
    not   & parameters in & from  & is when \\
    for   & using c\#  & is    & follow up \\
    javascript & windows phone & the   & code returned \\
    how   & when only & not   & a status \\
    server & what is & css   & sky in \\
    from  & to add & a     & object 2007 \\
    jquery & status code & javascript & model object \\
    %is    & source open & vs    & mobile phone \\
    %on    & sky in & of    & would cascade \\
    %studio & server sql & change & windows 8 \\
    \bottomrule
    \end{tabular}%
  \label{tab:edit-term-stat}%
\end{table}%
\textbf{Type of Query Edit}.
To explore further, we observe the type of edits users make during query modification. In Figure~\ref{fig:edit-add-delte-term} we see that for non-code search users often add or delete one term at a time to generate their edited query. On the other hand, with code-related search, most of the time users often achieve an edited query by adding or deleting two terms.

This phenomenon can be observed in Table~\ref{tab:edit-term-stat} which shows top unigram and bigram tokens added, as well as deleted terms found in our dataset for \textbf{code queries}. For instance, users often add "visual studio" to their query to clarify their search intent as something related to "visual studio" platform. Similarly, users often mention "using c\#" term to indicate they want the solution in "c\#" language. A similar conclusion can be drawn for the deleted term as well.

\begin{comment}
\begin{table}[htbp]
  \centering
  \caption{Changes in Codeness Score for Edited Code Query}
    \begin{tabular}{lcccc}
    \toprule
          &Inc. & NoChange . & Dec. & TotalEditQuery \\
    \midrule
    \#    & 16931 & 3054  & 10931 & 30916\\
    \%    & 54.76 & 9.88 & 35.36 & 100\\
    \bottomrule
    \end{tabular}%
  \label{tab:addlabel}%
\end{table}%

\end{comment}

\begin{figure}[!htpb]
\centering
\scriptsize
 \includegraphics[width=0.9\columnwidth]{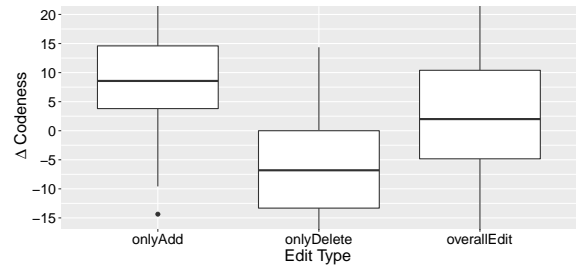}
 \caption{\textbf{\small Query Edit Type vs Codeness}}
 \label{fig:edit-vs-delta-codeness}
 \vspace{-0.1cm}
\end{figure}

\textbf{Edit vs Codeness Score}.
Figure~\ref{fig:edit-vs-delta-codeness} shows how \cscore changes when uses reformulate code related queries. We see that the developer achieve an edit by only adding terms, it often increases (\ie positive median of \textit{onlyAdd} in Fig.~\ref{fig:edit-vs-delta-codeness}) code intent of the query. Not surprisingly, when developers modify a query by only deleting terms, \cscore decreases (\ie negative median of \textit{onlyDelete} in Fig.~\ref{fig:edit-vs-delta-codeness}). However, we see that the median of \textit{overallEdit} is $2$ (\ie positive). This indicates, when the developer reformulates (\ie adding or/and deleting terms) a query, it often increases the code intent of the current query.

%Figure~\ref{fig:edit-vs-codeness} shows how different \cscore change (increase, decrease, and no change) in different codeness thresholds. We see that the \textbf{percentage} of edited queries increases with the increase of codeness value. This indicates a higher code intent query is more likely to be edited. In addition to that, as the codeness score increases, users are more likely to insert code-tokens than to delete them, which is indicated by "increase" and "decrease" lines in Figure~\ref{fig:edit-vs-codeness}. Finally, "nochange" edit remains similar throughout different codeness thresholds.

\begin{comment}
    \begin{figure}[!htpb]
    \centering
    \scriptsize
     \includegraphics[width=0.9\columnwidth]{Figures/edit-vs-codeness-plot.png}
     \caption{\textbf{\small Edit Query vs Codeness}}
     \label{fig:edit-vs-codeness}
     \vspace{-0.1cm}
    \end{figure}
\end{comment}

These results indicate that though in terms of time spending per query and website visit, there is no significant differences, users have to edit code queries more often than non-code queries. Thus, overall more effort is needed to query code using \GPSE.

% \RS{2}{Users 1) spend more time for code related search than non-code, 2) behave similarly in terms of number of result clicks, but the percentage of no-click is around $18\%$ less in code related search, and 3) often modify code queries to increase code intent.}

\RS{2}{Users modify code queries more often than non-code queries to retrieve desired results.}
%!Tex root=main.tex
\RQA{3}{ How do task sessions vary for code and non-code related search tasks?}
\label{rq3}

Often users need multiple queries to complete a task. 
Thus, we further check how much effort is needed to compare a whole code vs.~ non-code task. We use the annotated tasks data and perform the following experiments. 

\textbf{1) How many queries do users need to complete a search task?}

In Table~\ref{tab:query-per-task}, we see that the number of single query non-code task is $90.76\%$ which is significantly higher compared to $70.96\%$ for code task. On the other hand, the percentage of task consist of two query is $16.41\%$ for code which is significantly higher than $6.49\%$ for the non-code. As the number of queries per task increases ($2$, $3$, $4$, $5$, $5+$) (Table~\ref{tab:query-per-task}) the percentage of the task is getting higher for code task compared to a non-code task. Overall, code task requires significantly (Wilcox significance test with \textit{small} Cohen'D effect size) higher number of queries than non-code.
%When a user stops editing a query and issues a completely new query, this can be considered as a sign of user satisfaction~\todo{ref}. 
On the other hand, the number of queries to complete a task can be considered as the amount of effort and interaction required from users. This implies that the effort required to complete a search task is higher for code related search compared to non-code search. 

In addition, we see that (in Table~\ref{tab:query-per-task}) for code related search almost $2\%$ of the time user made more than $5$ edits to the queries which is around $85\%$ higher compared to non-code search (\ie $0.29\%$). This result indicates that users remain patient with the search engine when they look for the code. This also indicates a code search task is more complex which required more edits on queries to properly convey the information need to the search engine compared to non-code general search.

% Table generated by Excel2LaTeX from sheet 'task-query'
\begin{table}[t]
  \scriptsize
  \centering
  \caption{Number of Queries per Task}
        \begin{tabular}{ccc}
    \toprule
    \# Query & \% Code & \% Noncode \\
    \midrule
    1     & 70.96 & \textbf{90.76} \\
    2     & \textbf{16.41} & 6.49 \\
    3     & \textbf{6.22} & 1.55 \\
    4     & \textbf{2.94} & 0.67 \\
    5     & \textbf{1.51} & 0.24 \\
    5+    & \textbf{1.96} & 0.29 \\
    \midrule
    Total & 100\% & 100\% \\
    \bottomrule
    \end{tabular}%
  \label{tab:query-per-task}%
  \vspace{-3pt}
\end{table}%

\begin{comment}
   \begin{figure*}[!htpb]
    \centering
    \scriptsize
     \includegraphics[width=1.9\columnwidth]{Figures/query-per-task.eps}
     \caption{\textbf{\small Query/Task Statistics for Code vs Non-code}}
     \label{fig:query-per-task-stat}
     \vspace{-0.1cm}
    \end{figure*}
   
\end{comment}
    
\textbf{2) How much time do users need to complete a search task?}

\begin{figure}[!htpb]
    \centering
    \scriptsize
     \includegraphics[width=0.9\columnwidth]{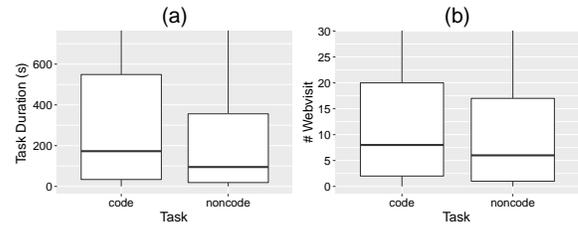}
     \caption{\textbf{\small Task Duration and Webvisit Statistics for Code vs Non-code}}
     \label{fig:task-duration-webvisit}
     \vspace{-0.1cm}
    \end{figure}

In this RQ we analyze how much time users spend on the search task. We sum up all the queries' activity time in a task to get the total duration of a search task. From Figure~\ref{fig:task-duration-webvisit} (a), we see that most of the code tasks required more time compared to non-code search tasks. The median time to complete a code search task is around $2$ min $53$ sec and the median time to complete a non-code search task is $1$ min $35$ sec. We see that generally, users spend almost twice the time for code related search compared to non-code. This result confirms the finding of RQ~\ref{rq3}-1 that users are more patient for code related search than for non-code related search.

\textbf{3) How many different website visits are required to complete a search task?}

Similar to time duration, we also analyze the number of web visits required for different tasks. In Figure~\ref{fig:task-duration-webvisit} (b), we see that most of the code search tasks required more web visits than non-code search tasks. In RQ~\ref{rq2}, we find no specific pattern in the number of web visits between code and non-code related query. However, here we see that the median of the number of web visits is $8$ for code task, which is higher than the median of the number of web visits for non-code task ($6$). 
The increasing number of queries in code search tasks (as in RQ~\ref{rq3}-1) might contribute to the difference in number of web visits for the search task.

%\textbf{4) How do users modify queries within a search task?~\todo{still in the research process. might remove if no convincing approach found}}

%\RS{3}{Users 1) issue a higher number of queries to complete a code task compared to non-code task and exhibit more patience when they look for code task, 2) spend almost double the time for code related search task compared to non-code, 3) visit more web pages to complete a code search task compared to a non-code search task}    
\RS{3}{Users spend significantly more effort for code related task than non-code related task in terms of number of queries, task completion time, and number of website visit.}

\subsection*{Discussion}

In summary, we find that code and non-code queries have different query characteristics. Also, user needs to put more effort to retrieve the intended results for code than non-code with a \GPSE. 

While our work primarily analyze code vs.~non-code queries, there could be further refinement possible for different kinds of programming tasks and information needs behind the search. For example, analyzing developers search queries on web, Xia \etal ~\cite{Xia2017} identified search tasks into seven different categories. We also see similar pattern while manually annotating $178$ code queries. Some sample queries with their task categories are shown in Table~\ref{tab:samle-query-multi-class}. 

% Table generated by Excel2LaTeX from sheet 'multi-code-class-sample'
\begin{table}[htbp]
 \scriptsize
% \small
  \centering
  \caption{Sample code queries with their task categories (from Dataset)}
    \begin{tabular}{cp{9.785em}p{14.285em}}
    \toprule
          & Task Type (short name) & Query Example \\
    \midrule
    1     & General Search (gen) & c\# property naming guidelines \\
    2     & Debugging (debug) & jira not loading images, Attempt to load Oracle client libraries threw BadImageFormatException This problem will occur ... \\
    3     & Programming (prog) & how to call class function in webservice c\# \\
    4     & Code Reuse (reuse) & GWTP template maven \\
    5     & Tools (tool) & php online debugger, lighttable \\
    6     & Database (db) & sqlserver database rename \\
    7     & Testing (test) & get protected member unit testing \\
    \bottomrule
    \end{tabular}%
  \label{tab:samle-query-multi-class}%
\end{table}%

\begin{figure*}[t]
\centering
\scriptsize
 \includegraphics[width=1.9\columnwidth]{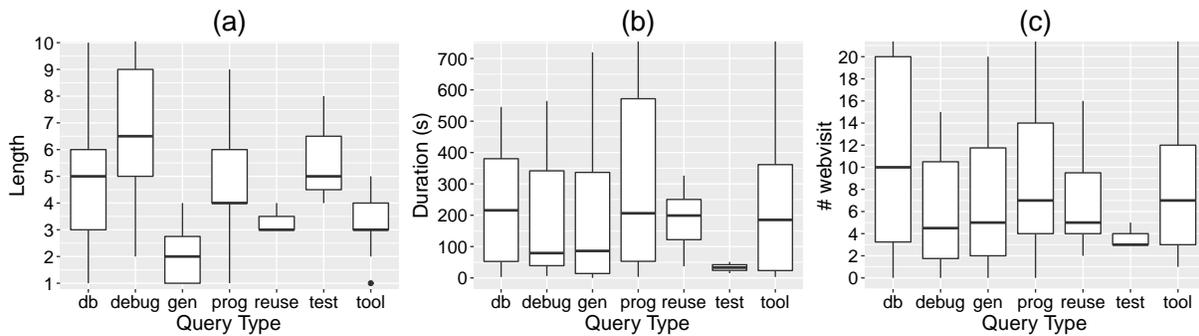}
 \caption{\textbf{\small Properties for different types of code queries}}
 \label{fig:multi-code-class}
 \vspace{-0.1cm}
\end{figure*}

Figure~\ref{fig:multi-code-class} further shows how query length, query browsing duration, and website visit can vary for different code tasks. 
We see that most of the \textit{debug} queries' length are higher compared to others and general code queries (\ie \textit{gen}) often smaller in length (in Figure~\ref{fig:multi-code-class} (a)). Sometimes, developers directly copy error message  and search with that in search engine (see debug query example in Table~\ref{tab:samle-query-multi-class}). This type of query are too specific and often intended for few web documents (\eg SO post discussion if exist). Thus, for \textit{debug} query developers often search for smaller duration (smaller median in Fig.~\ref{fig:multi-code-class} (b)) and visit smaller number of websites (smaller median Fig.~\ref{fig:multi-code-class} (c)).
In contrast, general code query (\ie \textit{gen}) often required lesser time - Fig.~\ref{fig:multi-code-class} (b)  and web visit - Fig.~\ref{fig:multi-code-class} (c). 
In other word, among code queries, {\sc GPSE}s are better at locating general code issues compared to other types (\ie debug, testing , etc.). We plan to do a detailed analysis for such different type of code queries in future.

%!Tex root=main.tex
\section{Study Implication}
\label{sec:dicussion}

\subsection{Implication of Code Intent Analysis}

Here we discuss some areas where the code intent analysis can be leveraged. %to classify code and non-code artifact, improve search engine system and generalization of the model.

\textbf{Search System}: It is important for the search engine to understand the search intent of users. Depending on the search intent returned results and other interaction with the search engine might vary. Search engines often use many meta information such as cookies, previous search history, URL click to understand users' search intent. 
This metasearch information is expensive to collect and not always available. For instance, the user might disable browser's cookie and history tracking or issue their first query. However, the model can predict a query intent on the fly and only requires the \textbf{query text}.
So, this code intent model can be used as a complementary tool which can be plugged into any existing search system. Further, search engine often suggests related queries to the issued (initial) query. 
Code intent model can be used to guide this query recommendation process. If the user shows initial code intent for a search, higher code intent queries can be suggested. 

%\noindent 
\textbf{Generalization}: To score a query or sentence, our model requires only a set of the domain-specific token ($S$). In this paper, to achieve code intent we leverage general code related token from Stack Overflow~\cite{stackoverflow}. This token set ($S$) can be easily extended or modified to identify specific code task. For example, "debugging" related tokens can be leveraged to predict whether the query intended for any code debugging task. Thus our model can be used to facilitate any further research where a fine granular code classification (\ie debug, testing, etc.) is required.
Additional knowledge about tags can be incorporated into the code intent analysis. For example, programming languages (\eg java, haskell) or related technologies (\eg react.js, mysql) can be assigned with constant scores. This process helps to mitigate any popularity bias of tags of the similar kind (\eg java, haskell).

In addition to predicting query intent, the model can be applied to score any document (\ie sequence of tokens). Developers can leverage \cscore to guide their document writing (\eg API documentation) to make it discoverable by the search engine for code related search.

%\noindent
% \textbf{Code vs.~Non-code Classification}:
% The model assigns a continuous value to the query. A threshold can be set to distinguish code query from non-code. Additionally, different \cscore ranges can be used to cluster different code queries with various code intent. Cluster with higher score contains queries with higher intent for code. In this paper, we empirically derive a threshold for \cscore to differentiate code queries from non-code.

\subsection{Implication of Empirical Findings}

Unlike general non-code search, code issues usually require much more consultation with different documents including text (\eg API documentation) and code (\eg source code, bug reports), as evident by our findings that developers need to query more to complete a task (RQ3). Thus, code search imposes unique challenges for search engines when they treat documents with mixture of code, and text similarly as general textual document (\eg news article). Thus, it is important for a search engine to incorporate effective retrieval models for code-mixed document and apply them effectively during code search.

We also report in RQ1 that code queries are more verbose and contain less vocabulary than non-code queries. Our code-intent model can be further trained with such characteristics, which can eventually impact \GPSE's search performance. For example, if code intention is known by \GPSE, it can restrict its search space. \GPSE can also leverage the frequent added and deleted terms in advance from code queries to anticipate users' intent and recommend related queries accordingly.

%!Tex root=main.tex

\section{Related Work}
\label{sec:related}

%\textbf{How Well Do Search Engines Support Code Retrieval on the Web?}
% In this section we discuss existing research works which are related to this paper.

There is substantial evidence in the literature to support the premise that developers use general purpose search engines during software development (\eg \cite{stolee2014tosem, sim2011well, hucka2016software, Xia2017, martie2017understanding}). 
Sim \etal~\cite{sim2011well} conducted a comparison study on various code search techniques of developers and observed that the general-purpose engine work better to find \textit{reference examples} than do dedicated code search tools. 
%components for as-is reuse and that participant obtained the best results using a general-purpose information retrieval site. 
Though code-specific search engines worked better in searches for subsystems, the general-purpose search engine, Google, worked better on searches for \textit{blocks of code}. Furthermore, code-specific search engines (\ie Koder and Krugle) perform better when searching for \textit{subsystems of code}~\cite{sim2011well}.
In a survey, Stolee \etal~\cite{stolee2014tosem} found that $85\%$ of developers search the web for source code at least weekly.
This result is echoed in a survey by Hucka \etal~\cite{hucka2016software}, which found that $93\%$ of the developers search on general-purpose web search engines for  "ready-to-use" software and $91\%$ of them search for source code. Additionally, Hucka \etal found that only $18\%$ of participant developers use specialized code search engines for source code search. 

Search logs of general-purpose search engines have been analyzed to identify different general query characteristics and users search behavior (\eg ~\cite{silverstein1999analysis}). 
Similarly, dedicated code search engines' logs have also been investigated to determine the use of code search engines, topics of search queries, and format of queries (\eg ~\cite{bajracharya2012analyzing, Brandt2009log, sadowski2015developers}).

In contrast, we analyze a search log of a general-purpose search engine (\ie Google) which consists of both code and non-code related searches. 
%We build a novel model which can identify code intent of queries and leverage that to differentiate code query from non-code query. 
We analyze query characteristics and users' search behavior for both code and non-query and explore the difference between them.
Additionally, we evaluate the performance of the search engine for code related search compared to non-code general search.

%We investigate developers' search behavior and performance of a general-purpose search engine, Google, for code related searches compared to non-code. Furthermore, we investigate the variety of queries and user behaviors in code vs non-code settings.

In a study on Google developers, Sadowski \etal~\cite{sadowski2015developers} observed that developers frequently search for code, conducting an average of five search sessions with a total of 12 queries in each workday. They also determined that programmers search to support a variety of information needs, such as looking for API examples, code understanding, debugging, or locating code snippets. The extensive use of code search engines in software development indicates that code search tools have a significant impact on developers' performance in practice ~\cite{sadowski2015developers}. 

In a large-scale survey, Xia \etal ~\cite{Xia2017} identified the frequency and difficulty of different software development related web search tasks. They classify frequent search tasks including exceptions/error handling, reusable code snippets, programming bugs, and third-party libraries. On the other hand, search tasks including performance, multi-threading, and security bugs, database optimization, and reusable code snippets are identified as most difficult search tasks. They also observed that developers are likely to use general-purpose search engines (\ie Google) to search for code.% or explanations of exceptions.

Martie \etal~\cite{martie2017understanding} found that iterative support in search engine can provide better experience on searching for some specific development tasks. They categorized developers' responses to a question about why they search and observed that developers often search for code to implement a feature, support a design decision, or meet problem specification. This work is complementary to previous studies in its focus on comparing and contrasting code-related and non-code-related search tasks.

\section{Threats to Validity}
\label{sec:threats}
\textbf{Internal Validity}.
%\textit{Code Intent.}
There might be some tokens (\eg newly published library and API name) that  carry code intent but are not included in our code token set. %This can happen due to the fact that nobody mentions that tag on SO. %For example, a newly published library and its related API name might not be found in the SO post immediately until someone posts about it in SO. 
This can lead to incorrect \cscore. % for such tokens. 
But developers often use some other tokens to describe such unknown terms  when they search. For example, in query "telerik raddataboundlistbox winrt", although the second token is not included in our tag list, our classifier still identifies it as code related query by leveraging other tokens ("telerik" and "winrt").

Our tags popularity measure would assign a higher score for the query "java iterate array" than "haskell iterate array", as the programming language tag "haskell" is less popular than "java" on SO.  Nevertheless, one could argue that the \cscore of both queries should be the same.~\todo{check the language} 
However, in our case, separating code from non-code queries, such scenario is unlikely to occur between a code and a non-code query~\todo{plural?}.
%\textit{Crawling: some page may have dynamic content, location-aware content, we could get actual content during those query URL were clicked}

%Other Search Constraints: In addition to the ranking algorithm, there are several other factors that might result in a poor performance of a search engine on a query. For instance, 

%\textit{Classification Error.}
We use an automatic classifier to separate code and non-code queries. However, the classifier might make mistake and that may impact our analysis. To mitigate this threat, we select a \cscore threshold where the classifier achieves a better trade-off of \textit{precision}, \textit{recall}, and \textit{F1-score} on manually annotated queries. 
However, due to the ambiguous nature of the query, it is nearly impractical to build a classifier which is absolutely accurate. Even we observe that the inter-annotator agreement is $0.85$ (\ie Cohen's Kappa Coefficient). 
%Interestingly, our human annotators' agreement is $93\%$ yet our automatic classifier achieves an F1-score of $87\%$ (details in RQ~\ref{rq1}). 
This indicates that our automatic classifier effectively resembles human annotators.

\textbf{External Validity}.
User's prior knowledge about a topic may impact the  search performance of search engines. For example, a senior programmer might have more knowledge about a certain development task and use a search engine to refresh their knowledge. In contrast, the scenario for a new developer might be different. In our search log dataset, we do not have access to users' details information except anonymous ID. So, we treat all users similarly. 

Further, we manually annotate $178$ code queries into further categories (\ie debug, general, etc.). This number of queries might not be adequate to come to a definite conclusion about their search characteristics. 
%We will continue our r research to further understand Our findings might shed light on further research in this line.~\todo{better wording} 

We only study Google search log. Other \GPSE may perform differently. However, since Google is the most popular \GPSE, we think our study can be representative of different code and non-code query behavior.

\textbf{Construct Validity}.
We use a search log which was collected using a  Google Chrome activity tracker plugin. One inherent limitation of such tracker is that it tracks all the browser activity regardless of whether it's a search activity or not. For example, there might be some cases where a user searches for something and promptly switch tab to visit other websites (\eg email, social media, etc.) and again comes back to the search activity. In such cases, some website might be incorrectly extracted as \textit{clicked URL} for a query. However, such occurrences are usually less in number and can happen for both code and non-code related searches. So our comparison study (\eg code vs non-code duration, number of web visit) would be less impacted by these threats.

%!Tex root=main.tex

\section{Conclusion}
\label{sec:con}
Developers often use general-purpose search engines which are usually not optimized for code related search. We explore whether such choice is optimal for code related search by analyzing a search log consisting of both code and non-code query. 
Firstly, we build an automatic classifier that identifies code and non-code query.

We find that query characteristics (\ie length) vary for code and non-code. We also observe that code related searching often requires more effort (\ie time, web visit, query modification, etc.) than general non-code search. %Furthermore,  to complete a code search task users require more effort, and users often show more patience when they look for code-related information.
We further discuss how our study can be leveraged to improve code search using general-purpose search engines.

\section*{Acknowledgements} 
This work is sponsored by the National Science Foundation (NSF) grant CCF-16-19123 and CNS-16-18771. 
The conclusions of the paper are of the authors and should not
be interpreted as necessarily representing the official policies or
endorsements, either expressed or implied, of NSF.

%\clearpage
\balance

\bibliographystyle{ACM-Reference-Format}
\bibliography{main} 

\end{document}